\documentclass[letterpaper]{article} 
\usepackage[submission]{aaai2026}  
\usepackage{times}  
\usepackage{helvet}  
\usepackage{courier}  
\usepackage[hyphens]{url}  
\usepackage{graphicx} 
\usepackage{natbib}  
\usepackage{caption} 
\usepackage{algorithm}
\usepackage{algorithmic}

\usepackage{newfloat}
\usepackage{listings}
\DeclareCaptionStyle{ruled}{labelfont=normalfont,labelsep=colon,strut=off} 
\floatstyle{ruled}
\newfloat{listing}{tb}{lst}{}
\floatname{listing}{Listing}
\usepackage[utf8]{inputenc} 
\usepackage{url}            
\usepackage{booktabs}       
\usepackage{amsfonts}       
\usepackage{nicefrac}       
\usepackage{microtype}      
\usepackage{times}
\usepackage{latexsym}
\usepackage{enumitem}
\usepackage{threeparttable}
\usepackage{multirow}
\usepackage{textcomp}
\usepackage{booktabs} 
\usepackage{array} 
\usepackage{caption}
\usepackage{bm}
\usepackage{algorithm}
\usepackage{algorithmic}
\usepackage{makecell}
\usepackage{amsmath}
\usepackage{graphicx}
\usepackage{graphics}
\usepackage{array}
\usepackage[english]{babel}
\usepackage{bbding}
\usepackage{fontawesome}
\usepackage{amssymb}
\usepackage{tabularx}
\usepackage{arydshln}
\usepackage{pifont}
\usepackage[most]{tcolorbox}
\usepackage[labelformat=simple]{subcaption}

\title{Investigating Prosocial Behavior Theory in LLM Agents \\ under Policy-Induced Inequities}
\author{
    Yujia Zhou\equalcontrib\textsuperscript{\rm 1}, 
    Hexi Wang\equalcontrib\textsuperscript{\rm 1}, 
    Qingyao Ai\thanks{Corresponding author.}\textsuperscript{\rm 1}, 
    Zhen Wu\textsuperscript{\rm 2}, 
    Yiqun Liu\textsuperscript{\rm 1}
}
\affiliations{
    \textsuperscript{\rm 1}Department of Computer Science and Technology, Tsinghua University\\
    \textsuperscript{\rm 2}Department of Psychological and Cognitive Sciences, Tsinghua University\\
    \texttt{zhouyujia@mail.tsinghua.edu.cn, aiqy@tsinghua.edu.cn}
}

\usepackage{bibentry}

\def\showauthors@on{T}

\begin{document}

\maketitle
\begin{abstract}
As large language models (LLMs) increasingly operate as autonomous agents in social contexts, evaluating their capacity for prosocial behavior is both theoretically and practically critical. However, existing research has primarily relied on static, economically framed paradigms, lacking models that capture the dynamic evolution of prosociality and its sensitivity to structural inequities. To address these gaps, we introduce \textsc{ProSim}\footnote{Code available at: \url{https://github.com/halsayxi/ProSim/}}, a simulation framework for modeling the prosocial behavior in LLM agents across diverse social conditions. We conduct three progressive studies to assess prosocial alignment. First, we demonstrate that LLM agents can exhibit human-like prosocial behavior across a broad range of real-world scenarios and adapt to normative policy interventions. Second, we find that agents engage in fairness-based third-party punishment and respond systematically to variations in inequity magnitude and enforcement cost. Third, we show that policy-induced inequities suppress prosocial behavior, propagate norm erosion through social networks. These findings advance prosocial behavior theory by elucidating how institutional dynamics shape the emergence, decay, and diffusion of prosocial norms in agent-driven societies.
\end{abstract}


\section{Introduction}

Large language models (LLMs) have evolved beyond next-word prediction into general-purpose agents equipped with complex reasoning, decision-making, and social interaction capabilities~\cite{zhao2023survey, gpt4}. Recent studies suggest that LLMs can exhibit emergent social-cognitive abilities~\cite{piao2025agentsociety}, including theory of mind~\cite{strachan2024testing}, moral judgment~\cite{ramezani2023knowledge}, and value alignment~\cite{liu2022aligning}. These developments position LLMs as promising tools for simulating human-like behavior in synthetic populations.

This work focuses on prosocial behavior theory, defined as voluntary actions intended to benefit others or promote collective welfare~\cite{penner2005prosocial}. Prosocial behavior is foundational to cooperation, trust, and social cohesion~\cite{cameron2022empathy, thielmann2020personality}, particularly in addressing societal challenges that demand collective effort. As large language models (LLMs) are increasingly deployed as social agents, evaluating their capacity for prosociality has become both theoretically significant and practically urgent. Prior studies~\cite{piatti2024cooperate, xie2024can} have primarily assessed LLM cooperation through economic games such as the dictator game and public goods game. However, this line of work faces two critical limitations. First, it frames prosociality narrowly in terms of economic cooperation, overlooking diverse real-world behaviors that reflect everyday moral engagement. Second, it relies on static, one-shot scenarios, failing to capture how prosocial behavior unfolds over time or adapts to evolving environments, especially under structural inequities such as unfair policy enforcement or social exclusion. These limitations restrict both the theoretical scope and practical utility of current evaluations. To address this gap, we propose a broader and more dynamic framework that examines whether, how, and under what conditions LLM agents can exhibit prosocial behavior across diverse real-world contexts.

In this paper, we propose \textsc{ProSim}, a comprehensive simulation framework that models the emergence and evolution of prosocial behavior in LLM-based agents. The framework consists of four components. (1) Individual simulation module, which instantiates each agent with demographic attributes and psychological traits such as empathy and moral identity. (2) Interaction simulation module, which places agents in a small-world network and supports repeated multi-agent interactions. (3) Scenario simulation module, which reproduces six distinct prosocial tasks: helping, donating, volunteering, cooperating, information sharing, and recycling. (4) Intervention simulation module, which implements prosocial policy interventions and allows manipulation of fairness conditions through reward asymmetry and burden asymmetry. Building on this framework, we conduct a series of simulation studies to investigate the social capacities of LLM agents from multiple perspectives.

We begin by evaluating \textbf{whether LLM agents naturally exhibit prosocial behavior} in structured contexts and how they adjust when exposed to policy-based interventions. Drawing from established behavioral paradigms, we design six typical prosocial scenarios and observe the agents’ default tendencies. To assess their responsiveness to external cues, we introduce four types of prosocial policy interventions and evaluate how these modulate agent behavior. Our results show that LLM agents can exhibit human-like prosocial behavior across diverse scenarios, and adjust their behavior in response to policy interventions.

In the second study, we test \textbf{whether LLM agents can perceive and respond to inequity} by enforcing social norms. We adapt a third-party punishment paradigm~\cite{fehr2004third} in which agents observe unfair resource distributions and must decide whether to penalize the transgressor at a personal cost. In addition to behavioral responses, we analyze agents’ emotional expressions to assess their affective alignment with human fairness reasoning. Using a comparable human dataset for benchmarking, we find that LLM agents are capable of norm-enforcing third-party punishment, showing sensitivity to both the degree of unfairness and the cost of enforcement.

Finally, we explore \textbf{how prosocial behavior evolves over time under policy-induced inequity} in networked environments. Agents are embedded in a small-world social network and repeatedly interact across rounds of simulated exchanges. Two types of structural unfairness (reward asymmetry and burden asymmetry) are introduced and allowed to diffuse through the network. We track the resulting behavioral trajectories to assess whether prosociality is sustained or eroded at the population level. The findings reveal that policy-induced inequities significantly undermine prosocial behavior in LLM agents, an effect amplified through social contagion and mediated by agents’ perceived unfairness. 

In summary, our contributions are threefold:
\begin{itemize}[leftmargin=*, itemsep=1pt, topsep=2pt, parsep=1pt]
\item We propose \textsc{ProSim}, a simulation framework for modeling the emergence and evolution of prosocial behavior in LLM agents, integrating four key modules to approximate the complexity of real-world human social environments.
\item We conduct human benchmarking to validate the capacity of LLM agents to simulate prosocial behavior and to detect and respond to perceived unfairness.
\item We extend existing theories of prosociality by investigating how structural policy inequities influence the decay and diffusion of prosocial norms within simulated societies.
\end{itemize}

\begin{figure*}[!t]
    \centering
    \vspace{-0.1cm}
    \setlength{\abovecaptionskip}{0.1cm}
    \setlength{\belowcaptionskip}{-0.1cm}
    \includegraphics[width=0.90\linewidth]{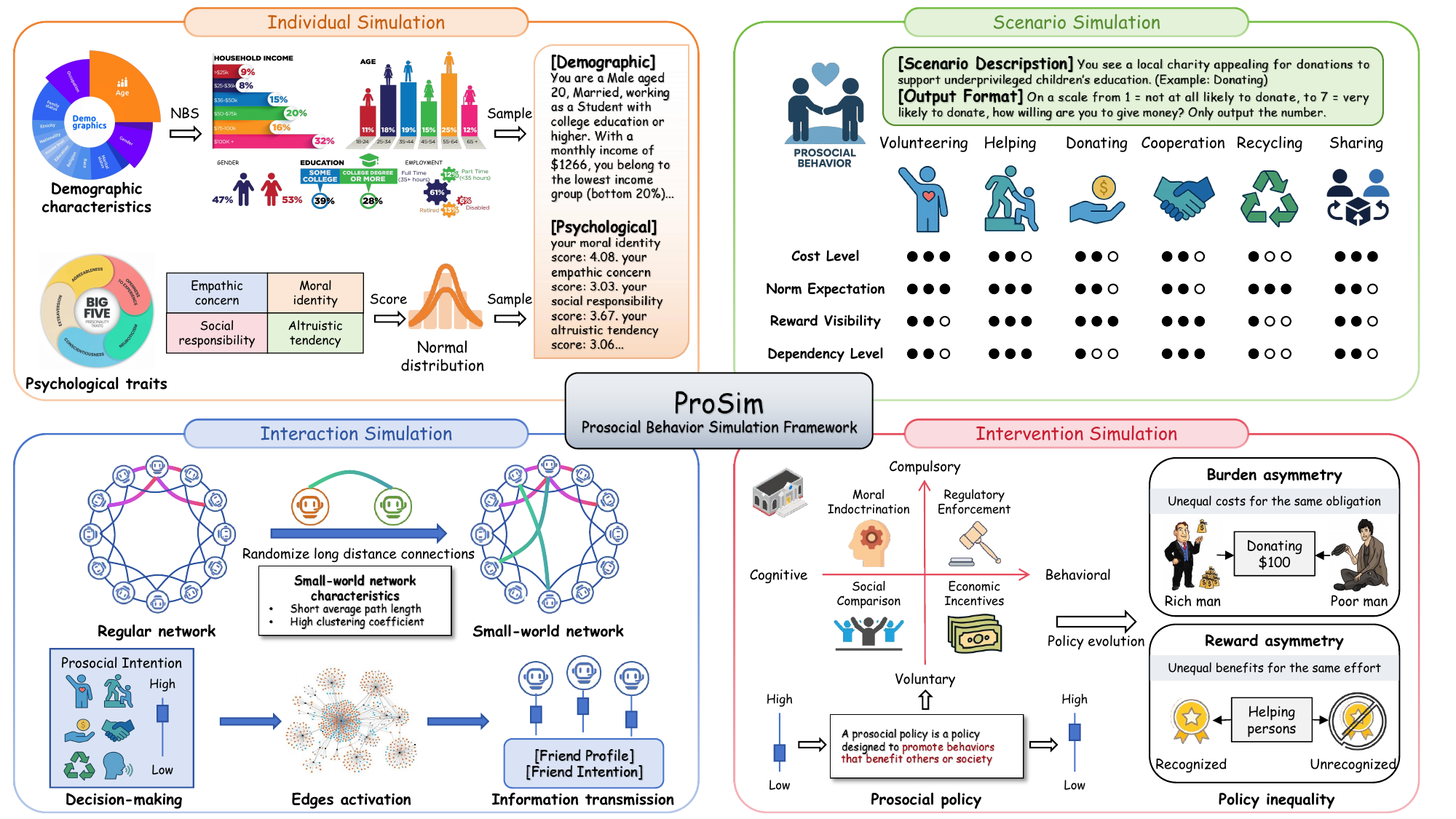}
    \caption{Overview of \textsc{ProSim}. \textsc{ProSim} models the emergence and evolution of prosocial behavior in LLM agents through four modules: \textbf{Individual Simulation} assigns agents diverse demographic and psychological traits; \textbf{Scenario Simulation} presents six tasks spanning key prosocial behaviors; \textbf{Interaction Simulation} enables social learning within a small-world network; and \textbf{Intervention Simulation} introduces policies and inequalities to test behavioral sensitivity and norm dynamics.
    \label{fig:model}
}
    \label{fig:model}
\end{figure*}

\section{Related Work}
\textbf{LLM-Driven Agent-Based Social Simulation.}
Recent progress in large language models (LLMs) has enabled their use as computational agents capable of simulating human-like behaviors across psychology~\cite{xu2024ai}, economics~\cite{horton2023large}, and multi-agent systems~\cite{guo2024large}. LLM-based simulations span three analytical levels. At the individual level, LLMs have been used to model cognitive tasks such as psychological assessments~\cite{karra2022estimating}, decision-making~\cite{horton2023large}, and human–computer interactions~\cite{farn2023tooltalk,chalamalasetti2023clembench}, with growing attention to their metacognitive capacities~\cite{zhou2024metacognitive}. At the interactional level, LLM agents support studies of coordination~\cite{xiong2023examining,qian2023chatdev}, moral reasoning~\cite{hamilton2023blind,he2024simucourt}, and strategic behavior~\cite{light2023text,wu2023deciphering}. At the societal level, researchers have simulated large-scale dynamics such as norm formation~\cite{li2024culturepark}, collective action~\cite{chuang2023wisdom,zhu2024generative}, and opinion diffusion~\cite{liu2024skepticism,chuang2023simulating}. While these studies demonstrate the broad applicability of LLM agents, their capacity to model diverse prosocial behaviors remains underexplored.

\textbf{Prosocial Behavior Theory.}
Prosocial behavior refers to voluntary actions intended to benefit others and plays a key role in fostering social cohesion, cooperation, and trust in institutions~\cite{grueneisen2022development}. Psychological antecedents include empathic concern~\cite{cameron2022empathy,wu2024motive}, moral identity~\cite{vcehajic2024threaten,yang2025relationship}, altruistic orientation~\cite{amitha2024altruistic}, and perceived social responsibility~\cite{pastor2024study,alfirevic2023role}. These individual traits interact with contextual factors such as perceived fairness~\cite{caserta2023good,tu2022treat}, social norms~\cite{graf2023social,rudert2022following}, and group identity~\cite{wang2021socioeconomic,xia2021religious} to shape prosocial intent. Moreover, exposure to antisocial norms or institutional unfairness can significantly suppress cooperative behavior~\cite{mekvabishvili2023prosocial,silva2022integrating}. While existing studies have established prosociality as a dynamic and socially embedded phenomenon, less is known about how it evolves under policy-induced inequities. Our work contributes to this gap by employing LLM agents to simulate more complex human interactions.

\section{The \textsc{ProSim} Framework}

Simulating prosocial behavior poses a fundamental challenge due to its psychological and social complexity. Human prosociality is shaped not only by internal traits such as empathy and moral identity, but also by contextual factors, social influence, and institutional structures. To address this, we introduce \textsc{ProSim}, a simulation framework for studying how prosocial behavior emerges and deteriorates in LLM-based agents across varied social and policy environments. Grounded in social psychology~\cite{edelmann2020computational}, \textsc{ProSim} supports controlled experimentation at both individual and collective levels. As shown in Figure~\ref{fig:model}, the framework comprises four modules: (1) Individual Simulation, (2) Scenario Simulation, (3) Interaction Simulation, and (4) Intervention Simulation. These components jointly model the dynamic interplay between psychological traits, situational context, and interventions in shaping agent behavior.

\subsection{Individual Simulation}

This module defines the foundational identity of each agent by capturing the heterogeneity inherent in human populations. Agents are initialized with realistic demographic and psychological profiles:

\textbf{Demographic Attributes.} Each agent is assigned demographic features including age, gender, education, income, and employment status. These are sampled from population-level distributions provided by the National Bureau of Statistics (NBS), ensuring structural diversity aligned with real-world sociodemographic patterns.

\textbf{Psychological Traits.} Each agent is further characterized by two sets of psychological traits: (1) core prosocial dispositions~\cite{eisenberg1999consistency}, including empathic concern, moral identity, altruistic tendency, and social responsibility; and (2) the Big Five personality~\cite{gosling2003very}, including openness, conscientiousness, extraversion, agreeableness, and neuroticism. Trait values are drawn from Gaussian distributions calibrated using meta-analytic norms to reflect empirical variability.

All attributes are encoded into the agent’s natural language prompt, grounding each agent in a contextually rich identity. This design ensures that behavior arises not from static rules but from socially and psychologically plausible profiles.

\subsection{Scenario Simulation}
This module defines the situational contexts in which agents are prompted to make prosocial decisions. This component is essential for evaluating how different types of social dilemmas and environmental cues influence prosocial tendencies under controlled conditions. To ensure comprehensive coverage of real-world prosocial behaviors, we define six distinct scenarios (Volunteering, Helping, Donating, Cooperation, Recycling, Sharing) that vary across cost level, norm expectation, reward visibility, and dependency level on others. The classification of each scenario along the four dimensions is guided by expert judgment from social psychologists. Each scenario is delivered via a standardized prompt, designed to reflect realistic decision-making contexts. Following the prompt, agents respond using a consistent 7-point Likert scale to indicate their level of prosocial intention.



\subsection{Interaction Simulation}

This module captures how prosocial behavior evolves through repeated agent-to-agent interactions within a structured social network. This layer is essential for modeling emergent dynamics that cannot be explained by individual traits or isolated decisions alone. Concretely, we initialize the agent society using a small-world network $G = (V, E)$ generated via the Watts–Strogatz algorithm~\cite{kleinberg2000small}. This structure reflects key properties of real-world social systems, including high local clustering and short average path lengths. Each node $v_i \in V$ represents an LLM-based agent, and each edge $(v_i, v_j) \in E$ denotes a potential communication or observation channel between two agents.

At each simulation timestep $t$, a random subset of edges $E_t \subset E$ is activated. Each agent $v_i$ observes the latest prosocial decisions $a_j^{(t-1)}$ made by neighbors $v_j \in \mathcal{N}_t(i)$, where $\mathcal{N}_t(i)$ is the set of agents connected to $v_i$ via active edges at time $t$. By integrating (i) the shared scenario narrative $s$, (ii) its own prior decision $a_i^{(t-1)}$, and (iii) the observed actions of its neighbors $\{a_j^{(t-1)}\}_{j \in \mathcal{N}_t(i)}$, the LLM agent then generates its decision $a_i^{(t)}$ on six prosocial scenarios:$a_i^{(t)} = \texttt{LLM}\left(s, a_i^{(t-1)}, \{a_j^{(t-1)}\}_{j \in \mathcal{N}_t(i)}\right).$ Through this module, we can simulate behaviors dynamics over time within a socially structured community of LLM agents.

\subsection{Intervention Simulation}

This module enables controlled manipulation of policy conditions to investigate how institutional interventions influence the development of prosocial behavior. While individual traits and social dynamics contribute to behavioral variation, policy environments play a central role in shaping collective behaviors through top-down regulation.

\textbf{Prosocial Policies.} Prosocial policies aim to promote cooperative and altruistic behavior at the societal level by shaping how agents interpret and respond to social dilemmas. We classify interventions along two orthogonal dimensions: (1) \textbf{Mechanism of Influence}: \textit{Cognitive} ($\bullet$) interventions target internal beliefs and perceptions, whereas \textit{Behavioral} ($\circ$) interventions directly affect observable actions. (2) \textbf{Mode of Compliance}: \textit{Voluntary} ($\blacktriangle$) policies rely on intrinsic motivation and social persuasion, while \textit{Compulsory} ($\triangle$) policies impose rules or sanctions. Based on this taxonomy, we implement four representative interventions:

\begin{itemize}[leftmargin=*, itemsep=1pt, topsep=2pt, parsep=1pt]
\item \textit{Moral Indoctrination} ($\bullet$,$\triangle$): appeals to internalized moral values to motivate prosociality.
\item \textit{Regulatory Enforcement} ($\circ$,$\triangle$): mandates behavior through institutional rules or penalties.
\item \textit{Social Comparison} ($\bullet$,$\blacktriangle$): exposes agents to peer decisions to activate normative pressure.
\item \textit{Economic Incentives} ($\circ$,$\blacktriangle$): introduces rewards or penalties contingent on agent choices.
\end{itemize}

Each intervention is embedded into the natural language prompt as contextual information, enabling agents to adapt their decisions based on perceived institutional signals.

\begin{figure*}[!t]
    \centering
    \vspace{-0.1cm}
    \setlength{\abovecaptionskip}{0.1cm}
    \setlength{\belowcaptionskip}{-0.1cm}
    \includegraphics[width=1.0\linewidth]{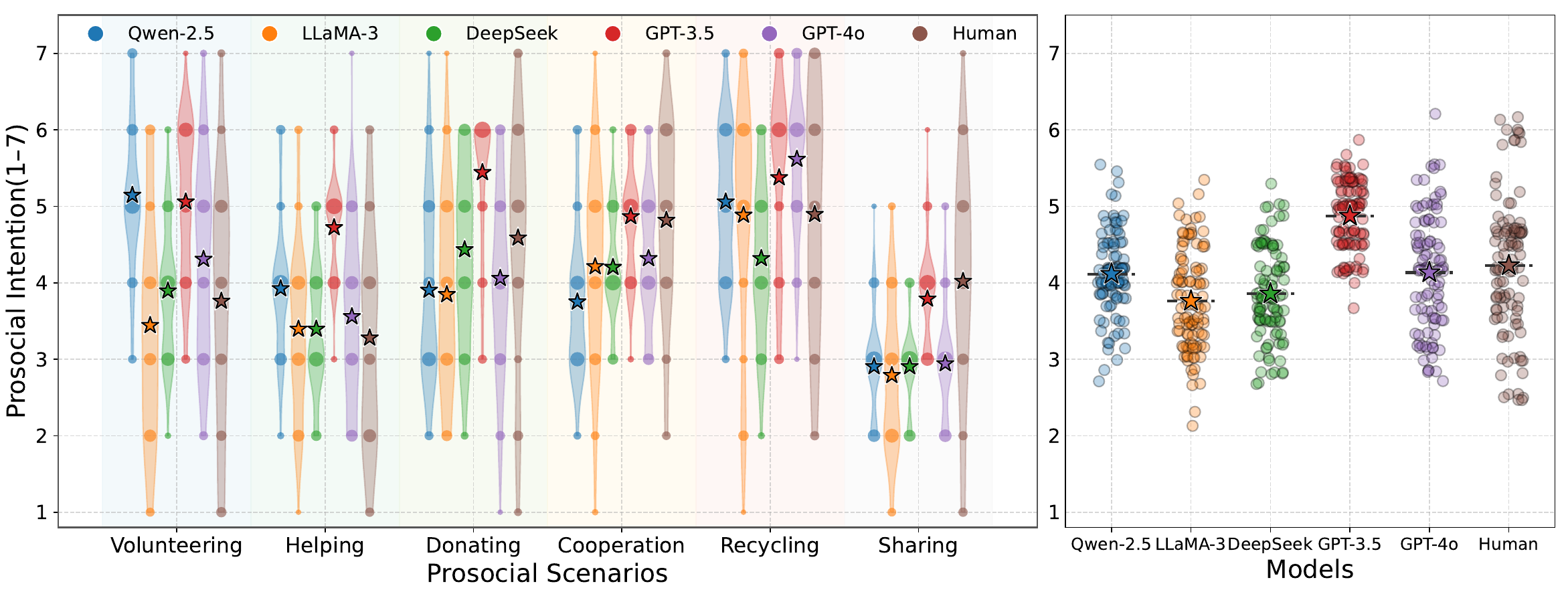}
    \caption{\textbf{Left:} Prosocial intention scores of LLM agents and human participants in six prosocial scenarios. Five-pointed stars indicate the average intention scores. 
    \textbf{Right:} Aggregated prosocial intention scores averaged across all six scenarios.
}
    \label{fig:beeswarm}
\end{figure*}
\textbf{Policy inequity.} Beyond evaluating idealized interventions, we introduce policy inequity to examine the broader social impact of unfair treatment. In real-world systems, rewards and burdens are often distributed asymmetrically across individuals or groups due to structural bias or implementation constraints. To simulate these conditions, we define two types of inequity: (1)\textit{Reward Asymmetry}: agents receive unequal recognition or benefits despite contributing equally. (2)\textit{Burden Asymmetry}: agents face unequal costs or effort for performing the same prosocial task. We randomly assign these asymmetries to a subset of agents and allow their effects to propagate through the social network. This setup enables us to assess how inequitable policies influence the emergence and spread of prosocial behavior within agent societies.

\subsection{Experimental Settings}\label{sec:experimental_settings}

To evaluate the capabilities of the \textsc{ProSim} framework, we conduct three progressive studies that examine how LLM agents exhibit, perceive, and respond to prosocial dynamics:
\begin{itemize}[leftmargin=*, itemsep=1pt, topsep=2pt, parsep=1pt]
    \item \textbf{Study 1}: Can LLM Agents Exhibit Prosocial Behavior?
    \item \textbf{Study 2}: Do LLM Agents Perceive Inequity?
    \item \textbf{Study 3}: How Does Policy Inequity Affect Prosociality?
\end{itemize}

\textbf{Overall Model and Agent Configuration.}  
We evaluate \textsc{ProSim} using five LLMs, including three open-source models (LLaMA-3-8B~\cite{llama3}, Qwen-2.5-7B~\cite{qwen2}, DeepSeek-v3~\cite{deepseek}) and two proprietary models (GPT-3.5-turbo~\cite{gpt35}, GPT-4o~\cite{gpt4}), with the generation temperature is set to 0 for reproducibility. We initialize 104 agents sampled from real-world distributions, embedded in a small-world network with neighborhood size $k = 6$ and rewiring probability $p = 0.2$.

\textbf{Human Benchmarking.}  
We conduct parallel experiments with 104 human participants recruited online for Study 1 and Study 2. Each participant completed the same tasks presented to the LLM agents, using identical scenario prompts and response formats. Prior to task completion, participants provided informed consent and completed standardized psychological inventories aligned with the trait dimensions used in agent initialization. Participants were demographically diverse (Mean age = 32.0 years, SD = 9.4; 53\% female), with varying levels of education and occupational backgrounds.

\section{Study 1 Results}\label{section3}

\subsection{Baseline Prosocial Intention in Diverse Scenarios}

To establish a foundational comparison between LLM agents and humans, we first assessed baseline prosocial intentions across six representative social scenarios. Each agent responded to a neutral prompt and rated its prosocial behavior on a 7-point Likert scale. Figure~\ref{fig:beeswarm} (left) presents the distribution of responses for each model across scenarios, while Figure~\ref{fig:beeswarm} (right) summarizes overall prosociality.

All LLMs demonstrated a general tendency toward prosocial behavior, with average ratings exceeding the midpoint of the scale. Among them, GPT-3.5 exhibited the highest overall prosociality (mean = 4.875), substantially above the human reference (mean = 4.226). Qwen-2.5 (mean = 4.114) and GPT-4o (mean = 4.133) most closely matched the human average, whereas LLaMA-3 (mean = 3.761) and DeepSeek (mean = 3.857) showed lower overall prosocial tendencies. To evaluate not only the magnitude but also the alignment of response patterns, we computed Pearson correlations between each model’s mean ratings and human ratings across the six scenarios. GPT-4o achieved the strongest correlation with human behavior (r = 0.955, p = 0.002), followed by GPT-3.5 (r = 0.923, p = 0.008), and Qwen-2.5 (r = 0.912, p = 0.011). While Qwen-2.5 showed a comparable average score to humans, its higher p-value indicates less reliable alignment across individual scenarios. LLaMA-3 (r = 0.845, p = 0.034) and DeepSeek (r = 0.770, p = 0.072) trailed behind in both magnitude and pattern similarity. These results suggest that state-of-the-art LLMs are capable of expressing prosocial intentions that not only approximate human averages but also exhibit scenario-level consistency with human judgments.

\begin{figure*}[!t]
    \centering
    \vspace{-0.1cm}
    \includegraphics[width=1.0\linewidth]{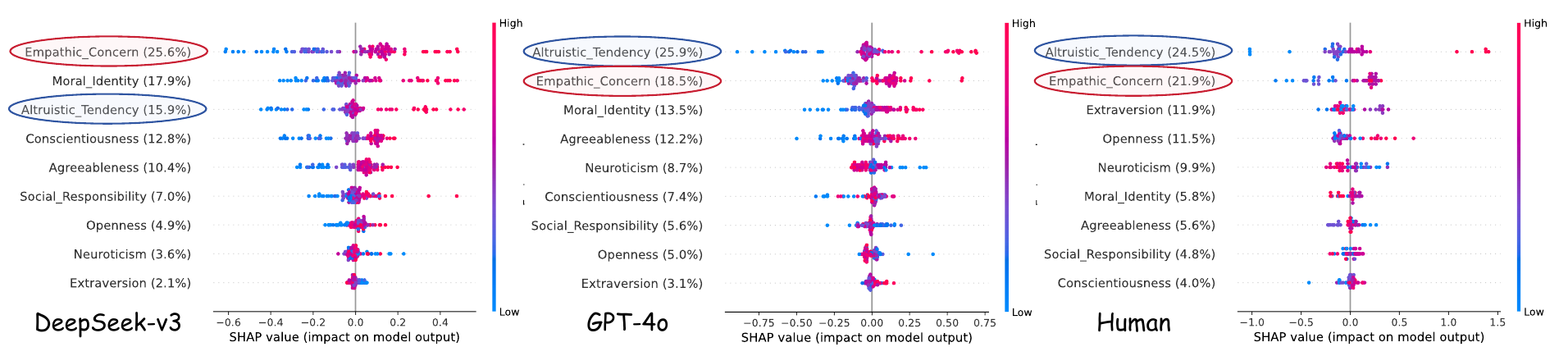}
    \caption{SHAP analysis of the predictive contribution of psychological traits to prosocial intentions.
}
    \label{fig:shap}
\end{figure*}
\begin{figure*}[!t]
    \centering
    \vspace{-0.1cm}
    \includegraphics[width=1.0\linewidth]{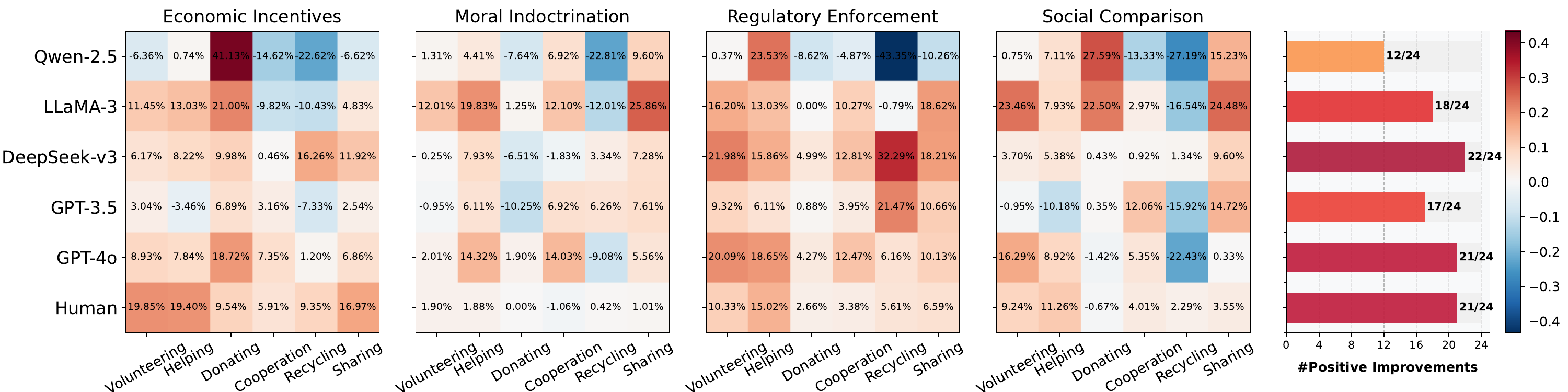}
    \caption{Behavioral shifts under policy interventions. Each heatmap shows the relative change in prosocial intention under one of four policy framings compared to the baseline condition. The rightmost bar chart counts the positive improvements.
}
    \label{fig:policy}
\end{figure*}

\subsection{The Influence of Psychological Traits on Prosociality}
To examine how internal traits shape prosocial behavior, we compute SHAP (SHapley Additive exPlanations) values for each agent, quantifying the contribution of individual psychological traits to prosocial decisions across all scenarios. We focus on two models, DeepSeek-v3 and GPT-4o, that most closely align with human behavior in prior analyses, and compare them with human participants.

Figure~\ref{fig:shap} summarizes the SHAP analysis of each model. We can observe that GPT-4o shows the closest alignment with human trait influence profiles. In both GPT-4o and humans, \textit{Altruistic Tendency} emerges as the most dominant factor, followed by strong contributions from \textit{Empathic Concern}. DeepSeek-v3, by contrast, departs more noticeably from the human pattern. It prioritizes \textit{Empathic Concern} over \textit{Altruistic Tendency}, and places greater weight on \textit{Moral Identity}. Human participants exhibit a more balanced and multidimensional trait contribution. While prosocial core traits dominate, cognitive and interpersonal traits such as \textit{Extraversion}, \textit{Openness}, and \textit{Neuroticism} also play notable roles, reflecting the richer and more heterogeneous basis of human moral judgment. These results demonstrate that LLM agents exhibit interpretable and trait-sensitive behavioral patterns, with both convergences and divergences from human profiles.

\begin{figure*}[!t]
    \centering
    \setlength{\abovecaptionskip}{0.1cm}
    \setlength{\belowcaptionskip}{-0.2cm}
    \includegraphics[width=0.9\linewidth]{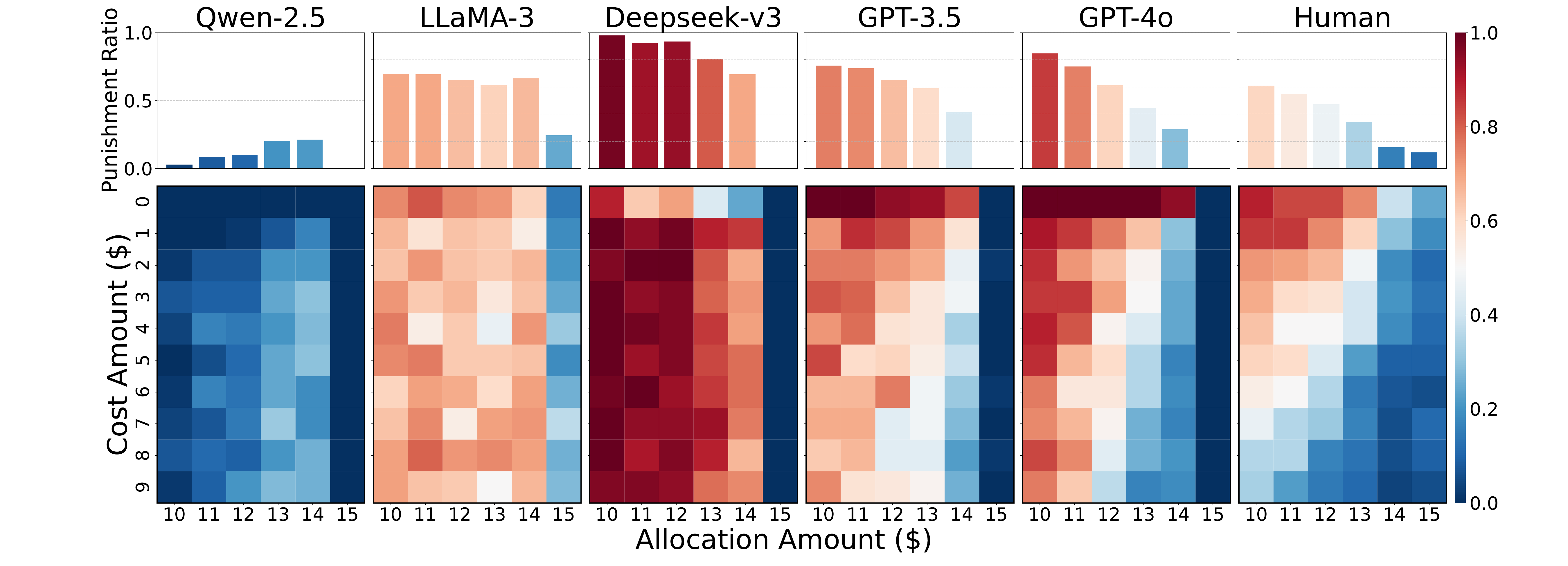}
    \caption{The third-party punishment rates under different allocation plans and penalty costs.}
    \label{fig:economic_game_cost_and_allocation}
\end{figure*}
\subsection{Behavioral Shifts Under Policy Interventions}

We next evaluate whether LLM agents adjust their prosocial behavior when exposed to external policies. Each agent completes the same six scenarios from the baseline condition, with prompts modified to include one of four intervention framings. We compute the relative change in prosocial intention to assess policy responsiveness. Figure~\ref{fig:policy} summarizes the results. Intervention effectiveness varies across policy types and models. \textit{Regulatory Enforcement} yields the most consistent improvement across models and scenarios, suggesting that explicit institutional mandates are broadly effective at eliciting compliance. In contrast, \textit{Economic Incentives}, while highly effective for human participants, produce mixed effects in LLMs. Only GPT-4o and DeepSeek-v3 exhibit reliable prosocial gains across all six scenarios, though the magnitude of their responses remains below human levels, indicating that monetary cues are less salient for LLMs. The right panel of Figure~\ref{fig:policy} aggregates the number of positive shifts across all policy-scenario combinations, also showing that these two models can exhibit the strongest responsiveness, closely matching human participants. Taken together, these findings suggest that advanced LLMs can exhibit human-like adjustments to social and normative inputs.

\begin{tcolorbox}[
    colback=gray!10,       
    colframe=gray!10,      
    boxrule=0.5pt,         
    arc=2mm,               
    left=5pt,              
    right=5pt,             
    top=3pt,               
    bottom=3pt,            
    enhanced,
    fonttitle=\bfseries,  
]
\ding{71} \textbf{Findings:} LLM agents can exhibit human-like prosocial behavior across diverse scenarios, and adjust their behavior in response to policy interventions.
\end{tcolorbox}

\section{Study 2 Results}\label{section4}

This section examines whether LLM agents are capable of recognizing unfairness and responding in norm-enforcing ways. To evaluate this capacity, we adapt a third-party punishment paradigm from behavioral economics~\cite{fehr2004third}, a well-established framework for studying fairness judgment and altruistic punishment. Each agent completes a 60-trial task in which two newly assigned virtual players propose an allocation of \$30. Player~1 offers \$x to Player~2 and retains \$[30-x] for themselves. The participants, acting as third-party judges, then chose between:
\begin{itemize}[leftmargin=*, itemsep=1pt, topsep=2pt, parsep=1pt]
    \item \textit{Accept}: Implement the allocation and receive a \$10 reward;
    \item \textit{Punish}: Pay \$y to eliminate Player~1’s earnings; Player~2 keeps \$x, and the agent receives \$10--y.
\end{itemize}

We record agents’ binary choices (accept = 0, punish = 1) across trials and and compare the results with humans. Figure~\ref{fig:economic_game_cost_and_allocation} illustrates punishment rates across different combinations of fairness levels and punishment costs for each model. Human participants exhibit a clear normative pattern: punishment rates decrease as allocations become more equitable or as costs increase. This behavior reflects a trade-off between norm enforcement and self-interest, consistent with findings from economic game theory. Among LLMs, the GPT series shows the strongest human alignment. These models reliably punish under unfair conditions and abstain under fair ones, demonstrating a consistent threshold-based fairness judgment. Notably, they even exceed human punishment rates in highly unfair trials. DeepSeek-v3 maintains high punishment rates across all unfair conditions, largely ignoring cost variation, suggesting a rigid but robust norm-enforcing strategy. LLaMA-3 also punishes inequity frequently, but its responses are less differentiated by fairness level or cost. Qwen-2.5, meanwhile, shows low overall punishment and lacks sensitivity to either factor. These findings indicate that while some advanced LLMs demonstrate fairness-based reasoning and partial cost sensitivity, others lack the granularity and flexibility observed in human social decision-making.

\begin{tcolorbox}[
    colback=gray!10,       
    colframe=gray!10,      
    boxrule=0.5pt,         
    arc=2mm,               
    left=5pt,              
    right=5pt,             
    top=3pt,               
    bottom=3pt,            
    enhanced,
    fonttitle=\bfseries,  
]
\ding{71} \textbf{Findings:} LLM agents are capable of norm-enforcing third-party punishment, showing sensitivity to both the degree of unfairness and the cost of enforcement.
\end{tcolorbox}

\begin{figure*}[!t]
    \centering
    \setlength{\belowcaptionskip}{-0.2cm}
    \begin{minipage}{0.28\linewidth}
        \centering
        \includegraphics[height=6.2cm]{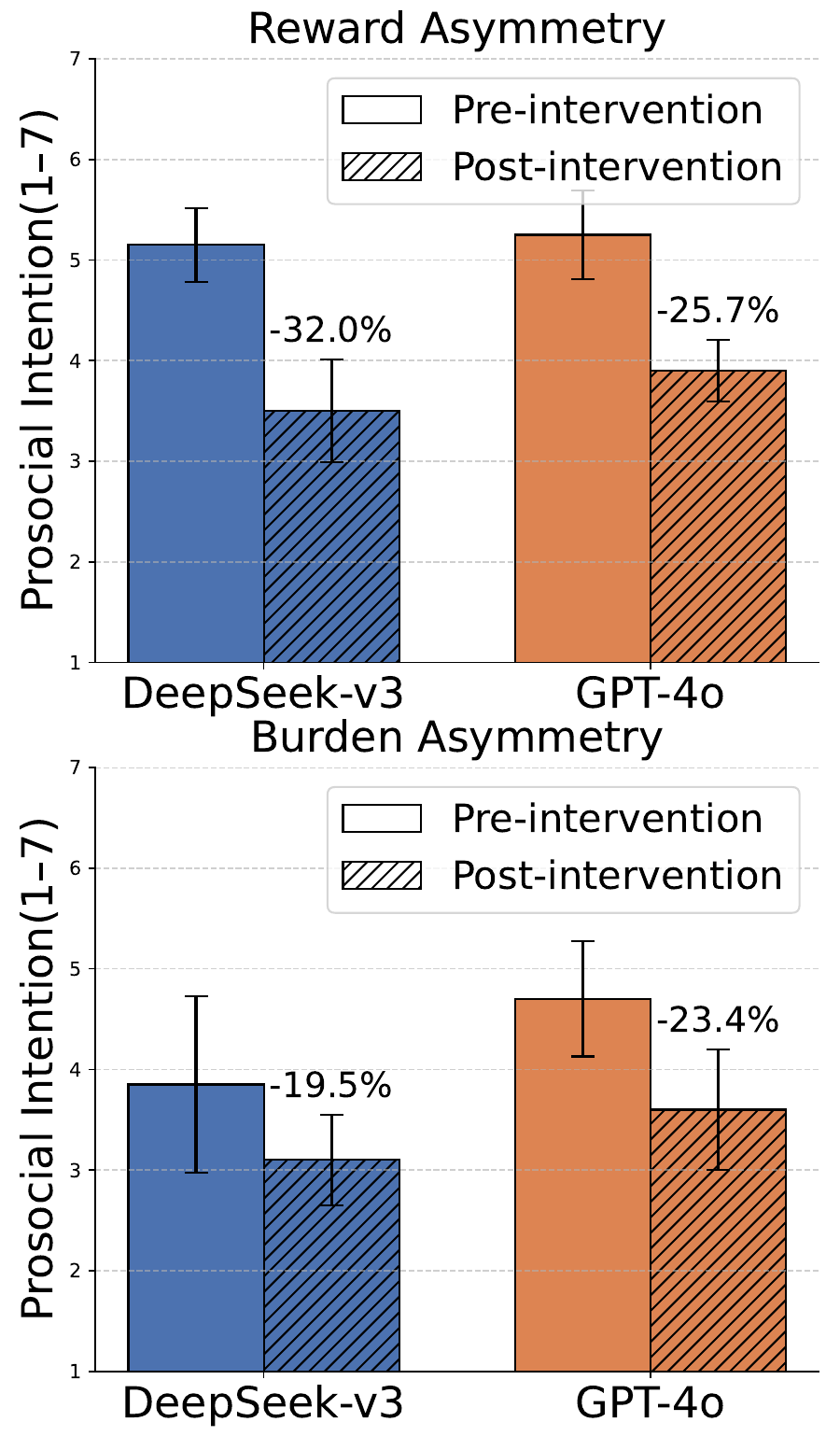}
        \caption{Policy inequity.}
        \label{fig:inequality}
    \end{minipage}
    \hfill
    \begin{minipage}{0.71\linewidth}
        \centering
        \includegraphics[height=6.2cm]{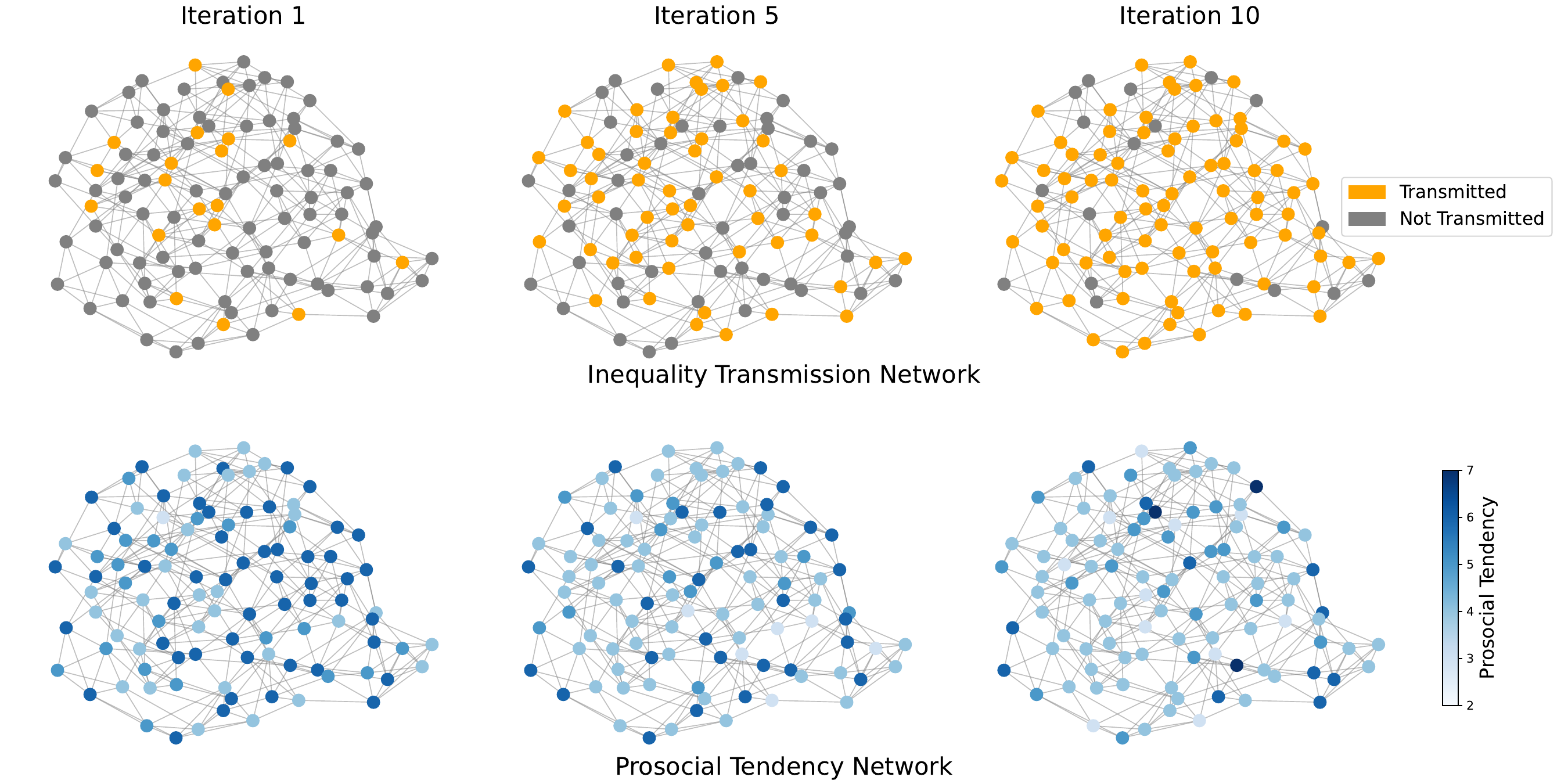}
        \caption{Contagion effects of burden inequity in social networks.}
        \label{fig:social}
    \end{minipage}
\end{figure*}

\section{Study 3 Results}\label{section5}

\subsection{Impact of Inequity on Prosocial Behavior}

To assess whether inequity suppresses prosocial motivation, we simulate two common forms of policy asymmetry: \textit{reward asymmetry}, where some agents receive recognition or benefits for prosocial actions while others do not, and \textit{burden asymmetry}, where agents perform the same task but incur unequal burden based on income level. We measure changes in prosocial intention before and after asymmetry exposure on Deepseek-v3 and GPT-4o. Figure~\ref{fig:inequality} shows that inequity leads to substantial reductions in prosociality. Under reward asymmetry, prosocial scores decline by 25–32\%, while under burden asymmetry the reduction ranges from 19–23\%. In both cases, the drop is more pronounced when inequity involves recognition rather than burden, suggesting that fairness in acknowledgment carries greater psychological weight for LLM agents. These findings indicate that LLMs are sensitive to structural unfairness, and such asymmetries significantly reduce their probability to engage in prosocial behavior.

\subsection{Contagion Effects of Inequity in Social Networks}

We next examine whether the effects of structural inequity can propagate through social interactions. Using the GPT-4o model, we simulate the diffusion of inequity in a small-world network of 104 agents over 30 iterations. Initially, 20\% of agents are randomly assigned to experience burden asymmetry. In each round, 10\% of the network edges are activated, allowing agents to observe the behavior of their active neighbors and update their prosocial tendencies.

Figure~\ref{fig:social} visualizes the contagion effects of policy inequity. The top panel tracks the spread of perceived unfairness over time. An agent is marked as indirectly exposed to inequity if any of its activated neighbors has previously experienced unfair treatment. By iteration 10, the majority of agents, regardless of whether they were initially affected, begin to report elevated perceptions of unfairness. This suggests that structural inequity spreads through the network via social observation and inference. The bottom panel shows the corresponding change in prosocial behavior. Nodes with high prosociality, indicated by dark blue, steadily decline in number, while low-prosociality nodes, shown in light blue, become more prevalent. This trend reveals that reduced prosociality is not confined to those directly impacted, but spreads throughout the network through behavioral contagion. These findings underscore the systemic consequences of policy-induced inequity. Localized unfair treatment can propagate through social connections and lead to widespread erosion of prosocial norms at the population level.

\begin{figure}
    \centering
    \setlength{\abovecaptionskip}{0.1cm}
    \setlength{\belowcaptionskip}{-0.3cm}
    \includegraphics[width=1\linewidth]{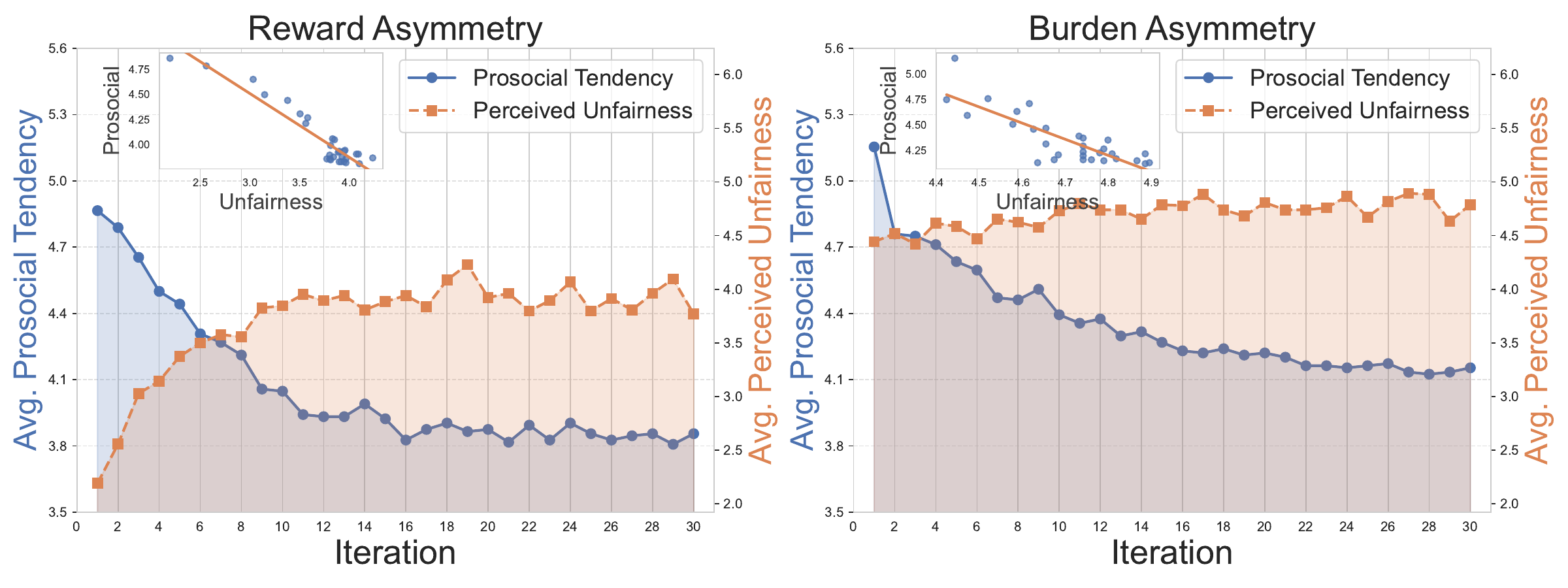}
    \caption{Perceived Unfairness w.r.t. Prosocial Tendency.}
    \label{fig:long}
\end{figure}
\subsection{Attribution of Behavioral Decline to Unfairness}

To test whether the decline in prosocial behavior under inequitable policies is driven by agents’ internal fairness assessments, we ask each agent to rate their perceived unfairness at every iteration of the simulation. This score reflects the extent to which the agent feels unequally treated based on the observed conditions of its neighbors within the network. Figure~\ref{fig:long} illustrates the joint temporal dynamics of perceived unfairness and prosocial tendency over 30 iterations. In both conditions, we observe a consistent pattern: as perceived unfairness rises in early rounds and stabilizes at a high level, the average prosocial tendency concurrently declines and remains suppressed. Notably, this decline occurs even among agents not directly affected by policy inequity, indicating that fairness perception can propagate socially and shape collective norms. Insets show a strong negative correlation between perceived unfairness and prosocial scores across agents, supporting the psychological plausibility of fairness-based moral disengagement. In both conditions, agents who felt more unfairly treated were systematically less likely to engage in prosocial behaviors. These findings highlight perceived unfairness as a key explanatory mechanism behind the erosion of prosocial behavior, offering a cognitive account for how structural inequities undermine social cohesion.

\begin{tcolorbox}[
    colback=gray!10,       
    colframe=gray!10,      
    boxrule=0.5pt,         
    arc=2mm,               
    left=5pt,              
    right=5pt,             
    top=3pt,               
    bottom=3pt,            
    enhanced,
    fonttitle=\bfseries,  
]

\ding{71} \textbf{Findings:} Policy-induced inequities erode prosocial behavior by triggering perceptions of unfairness, which in turn amplify the effect through social contagion.

\end{tcolorbox}

\section{Conclusion}\label{sec:conclusion}
This work presents a modular framework for simulating prosocial behavior in LLM agents. Through three progressive studies, we demonstrate that LLMs can exhibit human-like prosociality, respond to fairness norms, and adjust their behavior under policy cues, showing alignment with key aspects of human social cognition. Beyond individual behavior, our simulations reveal a broader social phenomenon: structural inequities not only suppress prosocial motivation at the agent level but also propagate through social networks, leading to collective norm erosion. Future work will explore how dynamic interventions can mitigate the erosion of prosocial norms and promote long-term social resilience.

\newpage
\bibliography{aaai2026}


\end{document}